# Econophysics
# and
# Financial–Economic Monitoring

## A. N. Panchenkov

*The Higher School of Economics – Nizhny Novgorod Branch*

*Russia, 603155, Nizhniy Novgorod, 25/12 Bolshaya Pechorskaya St.
e-mail: entropyworld@mail.ru*

The author solves two problems: formation of object of econophysics, creation of the general theory of financial-economic monitoring. In the first problem he studied two fundamental tasks: a choice of conceptual model and creation of axiomatic base. It is accepted, that the conceptual model of econophysics is a concrete definition of entropy conceptual model. Financial and economic monitoring is considered as monitoring of flows on entropy manifold of phase space – on a Diffusion field.

Contents
1. The deep reason of Econophysics occurrence
2. Results of Econophysics
3. Formation of object of Econophysics and creation of Axiomatic Base
4. The postulates of Econophysics
5. Entropy conceptual model of econophysics
6. Forecasting
7. Financial-economic monitoring: the definition
8. Phase space: global symmetry - the law of entropy preservation
9. Entropy manifolds: Hilbertian field
10. Diffusion: Diffusion field
11. Entropy Time
12. Financial-economic monitoring is the monitoring of flows on Diffusion field
13. The equations of the financial-economic monitoring
14. The matrix of density of an impulse
15. Destruction and restoration of financial - economic structures
    Bibliography

## 1. The deep reason of Econophysics occurrence

The time of econophysics occurrence is usually connected with the advent of the book by R. N. Mantegna and N. E. Stanley (2000) «Introduction in econophysics» [1]. The initial idea underlying creation of the new independent section of economy bases on an actuality of application of methodology, the theory and means of physics for research of developing economic systems, objects, processes etc. at their evolution.

This idea has led to widespread enough definition: *«econophysics is an interdisciplinary area of the researches using the theories and methods developed by physicists to solve problems of economy»*.

But here there is also a deep reason of econophysics occurrence.

The major discoveries of the second half of twentieth century have put forward a problem of transition from the concept and methodology of physics to more common basic entities and structures - concepts and methodology of Natural sciences. The fundamental fact has been found out - economy (as a science) is a part of Natural sciences. This event has led to revision of the bases of economy and in this act the generality of semantics, structure, behavior, characteristic properties and main laws of economy and physics has been established.

As a result we came to the deep reason of occurrence of econophysics: the economy, as well as physics, is the section of the common science - Natural sciences.

## 2. Results of Econophysics

Econophysics, being in the early period of the existence, develops actively and effectively; numerous teams of physicists, economists, mathematicians participate in its researches practically in all leading countries of the World. By now, some monographs and reviews devoted to various key sections and questions of econophysics [1-15] have been published and known.

So, it is possible already to speak about results of econophysics evolution.

First of all, we are interested in conceptually-methodological results. Let us list three conclusions, derived out of experience of econophysics evolution, formalized as the postulates:

**Postulate 1.**
*Conditions of financial-economic structures and objects are realized in chaos.*

**Postulate 2.**
*Financial-economic processes are the diffusion processes.*
**Postulate 3.**
*Financial-economic processes are the processes of self-organization, in which diffusion is controlled by the current chaotic process.*

## 3. Formation of the object of Econophysics and creation of the Axiomatic Base

The conceptual model of a separate science or Natural sciences as a whole rest upon two basic essences:
1. An object;
2. A fundamental extreme principle.

The object of a science should have axiomatic definition; the axiomatics of object defines conceptual integrity of the particular section of Natural sciences (particular science) as well as the level of a generality and natural-science importance of a new conceptual model.

All classical sections of Natural sciences (including sections of physics and mathematics) have the axiomatically defined objects of research, unlike the new sections of the science which have arisen in the second half of twentieth centuries.

Econophysics has yet no object, but the initial period of its development has not passed for nothing; there was an intensive search and gathering of the scientific facts forming its phenomenological base. Now the moment has come when the phenomenological base of econophysics contains necessary knowledge for formation of the object of econophysics.

The moment of the beginning of researches on creation of the object of econophysics has come.

The two actions play the key role in this fundamental problem:
1. A choice of conceptual model;
2. Creation of axiomatic base.

The major role of axiomatic base of econophysics in its further development and importance for a civilization are obliged to the fact that in the course of realization of the axiomatic approach the object of econophysics should be set by an axiomatic definition; and a generality, efficiency and correctness of the object of econophysics are completely determined by the axiomatic base.

## 4. The postulates of Econophysics

As economy as the science is the section of Natural sciences, let us list the three known postulates of Natural sciences [16]:
- A postulate of continuum;
- A postulate of duality;
- A postulate of duality of a condition.

The three postulates of Natural sciences have the following formulations:

**The postulate of continuum**

*All structures of the Universe and the surrounding Reality arise, function and collapse in a continuum.*

**The postulate of a duality**

*Fundamental symmetry of Natural sciences is the duality.*

**The postulate of a duality of a condition**

*Conditions of objects of the Universe and the surrounding Reality contain regular and singular components.*

To the three postulates of Natural sciences it is necessary to add **the principle of a limiting incorrectness**:

*Objects of the universe and the Validity surrounding us do not possess a limiting correctness.*

Three postulates of Natural sciences and principle of a limiting incorrectness define an ontological level where the conceptual model of econophysics exists and is realized.

The first general axiom plays the important role in the axiomatic base of econophysics.

**The axiom of chaos**

*Chaos - the organized medium, containing determined geometrical objects, entities and structures.*

The second general axiom has the formulation:

**The axiom of a condition**

*Two kinds of a condition exists in econophysics:*
1. *Movement;*
2. *Event.*

The structure of axiomatic base of econophysics includes also the three postulates of the paragraph 2.

## 5. The entropy conceptual model of Econophysics

The formulated postulates have found the first application at a choice of conceptual model of econophysics. In modern Natural sciences the entropy conceptual model developed by A. N. Panchenkov in 4 volumes of the monograph "Entropy" [16-19] has achieved the popularity and propagation.

The postulates of econophysics satisfy the entropy conceptual model. This fact has given the basis for the choice of the entropy conceptual model.

**The postulate of a choice** has the formulation:

*The conceptual model of econophysics is the concrete definition of the entropy conceptual model.*

The postulate of a choice puts a problem of formation of axiomatic base of econophysics on a real basis; first of all it concerns the object of econophysics. The object of econophysics is the concrete definition of the object of entropy.

This property provides the basis for the first special axiom of econophysics.

**The axiom 1.**
*The object of econophysics is the economic continuous medium.*
The corollary of the axiom 1 is
**The definition 1.**
*Econophysics is a science about conditions of financial - economic structures and objects in the economic continuous medium.*

The key moment exists here, which leads to the effective and constructive concrete definition of the object of econophysics.

According to the postulate 2 the following axiom is generated:
**The axiom 2.**
*The economic continuous medium is the diffusion continuous medium.*

Let us remind, that axiomatic definition of the virtual continuous medium is given in the book [16]. It is useful to note, that the experience of development of econophysics has led to the entropy conceptual model of econophysics. The philosophical interpretation of entropy in the new picture of the World is described in article [20]. The systematical account of various problems, tasks and theories of entropy and statistical physics is given in the books [21-42].

## 6. Forecasting

Forecasting in a broad sense is the central problem of econophysics.

For example, the basic problem in an estimation of risky financial assets is the prediction of the future on the basis of the present (current) information. Forecasting is one of the basic tools, methods, elements of economic planning and modeling. In the certain sense, studying of the dynamics of financial-economic objects, systems, structures and entities in chaotic diffusion environment is devoted to the purpose of forecasting. In this connection we shall note, that one of the key problems of econophysics is stochastic dynamics of stock exchange.

In this direction the three mathematical models have been created and received the greatest popularity:
1. Cox's, Ingersoll's and Ross's model [43,44];
2. Black's and Scholes's model [45];
3. Heston's model [8,46].

During the current period of econophysics evolution the problem of conceptualization and formalization (the formal description) of the task and the idea of forecasting has arisen and got the special importance.

In the conceptual model, methodology and theory of entropy this problem has the effective solution as the theory of monitoring. With reference to econophysics we mean the theory of financial - economic monitoring. We need a transition from the problem of forecasting to the problem of financial-economic monitoring.

## 7. Financial-Economic Monitoring: the definition

In econophysics financial-economic monitoring is introduced by the following definition:

**Definition 2.**

*Financial-economic monitoring is a supervision, determining and the analysis of a current condition located in the diffusion continuous medium of the financial-economic object in order to predict behavior of financial assets and structures, perform quantitative analysis and manage financial risks, investments and a capital, and to prevent crisis states and develop a strategy of restoration of an object structure.*

It is necessary to pay attention to the fact that the modern econophysics as the section of Natural sciences has yet no precise formalization whereas financial-economic monitoring has. Let us note, that so far the only theory of monitoring – the entropy theory of monitoring, described in books "Entropy" and "Entropy 2: Chaotic mechanics", - is known [16, 17].

It is also important that the problem of forecasting (especially, forecasting of a residual resource) is significant in mechanical engineering, energetics, manufacturing and operation of various kinds of machines. Therefore the theory of reliability and, first of all, the continual theory of reliability developed by V. V. Bolotin [47,48], should be considered as the progenitrix of the theory of financial-economic monitoring. But the problem of financial-economic monitoring should represent considerable generalization and expansion of the problem of reliability. In particular, the theory of reliability is the theory of deterioration of an object, this property is not applied to the theory of financial-economic monitoring. There should be an alternation of two cycles - development and ageing - in economical dynamics.

## 8. Phase space: global symmetry – the law of Entropy preservation

In the entropy model of econophysics and financial-economic monitoring the basic geometrical object is the phase space – the smooth manifold with local coordinates q and p :

$$\Omega = \{q, p \mid \Omega = \Omega_q \times \Omega_p; \Omega_q \subset R^n; \Omega_p \subset R_n; \Omega \subset R^n \oplus R_n \} .$$

Here:
q - the generalized coordinate;
p - an impulse;
$R^n$ - n-dimensional material Euclidean space;
$R_n$ - connected n-dimensional material Euclidean space.
The phase space forms:
1. Configuration space

$$\Omega_q = \{q \mid \Omega_q \subset R^n\}.$$

2. Space of an impulse

$$\Omega_p = \{p \mid \Omega_p \subset R_n\}.$$

General entropy has dual representation:

$$H_f = H_q + H_p; \{q,p\} \in \Omega. \qquad (1)$$

Here:

$H_q$ – structural entropy,

$H_p$ – entropy impulse.

The global symmetry

$$H_f = const. \qquad (2)$$

is supported on conditions of medium, located in phase space.

This symmetry is corollary from application of the principle of entropy maximum to Boltzmann's representation of entropy [21]

$$H_f = -\int_\Omega \rho \ln \rho \cdot d\Omega, \{q,p\} \in \Omega. \qquad (3)$$

In this initial representation of entropy $\rho$ - density of the virtual continuous medium.

## 9. Entropy manifolds: Hilbertian field

The first narrowing of phase space - the smooth manifold, called entropy manifold - is organized with the help of symmetry (2)

$$Э = \{q, p \mid Э \subset \Omega, H_f\}. \qquad (4)$$

The entropy manifold has a structure of the direct product:

$$\begin{aligned} Э &= Э_q \times Э_p; \\ Э_q &= \{q \mid Э_q \subset Э, H_q\}; \\ Э_p &= \{p \mid Э_p \subset Э, H_p\}. \end{aligned} \qquad (5)$$

Here:
$Э_q$ – the entropy manifold of configuration space,
$Э_p$ – the entropy manifold of impulse space.
The solenoid manifold is produced by setting the divergence on the entropy manifold

$$\sigma = \mathrm{div}A; \quad A = \left[\frac{\partial q}{\partial t}, \frac{\partial p}{\partial t}\right];$$
$$\mu = \{q, p \mid \mathcal{M} \subset \mathcal{Э}, \sigma = \mathrm{div}A\}. \tag{6}$$

On the solenoid manifold the symmetry (2) is supported by the equation

$$\sigma = 0; \quad \{q, p\} \in \mu \tag{7}$$

Introducement of the potential of accelerations in the theory results in the next narrowing of the entropy manifold to the manifold of potential of accelerations

$$\pi = \left\{q, p \mid \pi \subset \mu\,;\, \Theta\,;\, \xi = \begin{pmatrix} 0 & -E \\ E & 0 \end{pmatrix}\right\}. \tag{8}$$

Here:
$\Theta$ - potential of accelerations;
$\xi = \begin{pmatrix} 0 & -E \\ E & 0 \end{pmatrix}$ - alternate matrix of metric (metrics),
E - unity diagonal matrix.

As is known, the canonical equations of the potential of accelerations are held on the manifold of potential of accelerations [16].

$$\begin{aligned}\frac{\partial q}{\partial t} &= -\frac{\partial \Theta}{\partial p}; \\ \frac{\partial p}{\partial t} &= \frac{\partial \Theta}{\partial q}\end{aligned} \quad ; \quad \{q, p\} \in \pi. \tag{9}$$

The narrowing of the the potential of accelerations, containing the potential of impulse, occupies one of the central places.

The submanifold of the potential of acceleration, containing the potential of an impulse, is called the Hilbertian field.

The Hilbertian field looks like:

$$\Gamma = \{q, p \mid \Gamma \subset \pi; \Psi\}. \tag{10}$$

On the Hilbertian field

$$p = \mathrm{grad}\,\Psi;\ \Psi(q,t)$$

and the equation of the potential of accelerations [16, 18, 35] is true

$$\frac{\partial \Psi}{\partial t} = \Theta\ ;\ \{q,p\} \in \Gamma. \qquad (11)$$

### 10. Diffusion: Diffusion field

The diffusion plays the determining role in the methodology and the theory of econophysics and financial-economic monitoring; so, we have need for the entropy manifolds, supporting flows of the diffusion continuous medium. Setting of the potential of accelerations as:

$$\Theta = -(\dot q \mid p)_{R^n} - \Pi \qquad (12)$$

allocates the submanifold on the Hilbertian field

$$L_s = \{q, p \mid L_s \subset \Gamma; \Theta = -(\dot q \mid p)_{R^n} - \Pi\}. \qquad (13)$$

The basic geometrical object of econophysics is designed as narrowing of manifold $L_s$ by means of setting the two known fundamental entropy structures on this manifold:

$$\Psi = H_q;\ H_q = -\ln \eta. \qquad (14)$$

As a result, the new manifold, called «Diffusion field», is formed

$$D_s = \{q, p \mid D_s \subset L_s; \Psi = H_q; H_q = -\ln \eta\}. \qquad (15)$$

Here $\eta$ - density of the diffusion continuous medium.

The Diffusion field is the base geometrical object for the entropy description of the diffusion processes. The distinctive feature of Diffusion field is that basic equations of Hilbertian field on it allow the adequate formulation in terms of the structural entropy and density of the virtual continuous medium of configurational space.

The second extremely important fact: *the freedom of a choice of a vector field* $\dot{q}$ *of the manifold* $L_s$ *on Diffusion field was preserved.* This freedom is necessary for the full realization of the idea of description of economic casual processes using tools of the theory of entropy.

Here the situation is designed, corresponding, to a certain extent, to known (in theory of stochastic processes) situational duality «the stochastic differential equations – Fokker - Planck equations » [49, 50] .

The two basic sections, determining the base of all Natural sciences, exist in the physics of XX century. They are classical mechanics and the theory of diffusion. The classical mechanics underlies the description of all entities, having materialization and where the concepts of kinetic and potential energy are reasonable. The theory of diffusion is closely connected to heat conductivity and thermal energy. It has been established by scientists for a quite long time, that laws of heat conductivity are not laws of classical mechanics, which fact gave rise to the extremely important and interesting problem of interrelation between the theory of diffusion and classical mechanics.

This problem has been discussed regularly and sometimes actively in XX century. Now the problem of the joint analysis of classical mechanics and diffusion must be solved. Otherwise, it is hard to expect considerable progress in econophysics. A full clearness should be here, because the problem of the joint analysis itself should be one of the primary tasks of the general theory of econophysics.

## 11. Entropy Time

The brief review of the problem of time in econophysics we shall begin with a metaphor prevailing in Natural sciences «Time is a key to knowledge of the Nature». The problem plays a key role in econophysics as well.

The two times exist in the entropy conceptual model of Natural sciences:
1. Astronomical time t.
2. Entropy Time S.

The parametrization of conditions of flows on entropy manifolds is performed by means of astronomical time; whereas Entropy Time is the integral element of any object suitable for the entropy description. Here Entropy Time means internal (own) time [17, 18, 19]

The two times are introduced in econophysics by the postulate:

**The postulate of time**
*The two times exist on Diffusion field:*

*1. Astronomical time* t
*2. Entropy Time* S.

The fundamental fact lays in a basis of the entropy theory of internal time:
*Structural entropy is Entropy Time:*

$$S = H_q.$$

The equation of Entropy Time follows from the known equation of structural entropy

$$\frac{dH_q}{dt} = \sigma_1; q = D_S, t \in [0, T]. \tag{16}$$

Here:

$\sigma_1$ - divergent invariant

T - time of destruction.

The basic importance of Entropy Time in the theory of financial-economic monitoring is determined by the fact that at the instance of monitoring organization we need the measure of the current condition of object of monitoring. Entropy Time acts as this measure.

**The Postulate of the measure**

*Entropy Time is the measure of the current condition of an object of monitoring in the theory of financial-economic monitoring.*

## 12. Financial-economic monitoring is the monitoring of flows on Diffusion field

The vector of the derivative of generalized coordinate in the potential of accelerations (12) defines a flow on Diffusion field:

$$\dot{q} = u; q = D_s; t \in [0, T]; u = u(q, t) \tag{17}$$

The vector nonlinear ordinary differential equation (17) is the equation of a flow. This equation describes dynamics of financial-economic structures and belongs to the basic equations of the monitoring.

Let's address now to an impulse.

The duality of representation of an impulse - the second symmetry - plays the important role in econophysics:

$$p = \begin{cases} p \in \Omega_p \\ p(q,t); \ q \in \Omega_q; \ t \in [0,T] \end{cases} \quad (18)$$

The first component of this duality defines the free impulse, and the second component defines the attached impulse. In particular, the attached impulse is realized in case of diffeomorphism:

$$T_S : \Omega_q \to \Omega_p \cdot p(q,t) \in C^\infty(\Omega_q^T); \ \Omega_q^T = \Omega_q \times [0,T]$$

Let us note now the key moment: on the Diffusion field exists only the attached impulse. At conceptual registration this result is formulated as the special axiom of econophysics.

**The axiom of an impulse**
*there is no free impulse in econophysics.*

This axiom has deep sense: the economic continuous medium is the massless medium which does not contain objects and structures with final mass.

## 13. The equations of financial-economic monitoring

The mathematical model of the financial-economic monitoring includes the following complete set of the basic equations:

1. The equation of a flow

$$\dot{q} = u; \ q \in D_S; \ t \in [0,T]; \ u = u(q,p,t); \ p = p(q,t)$$
$$p \in C^\infty(D_S^T); \ D_S^T = D_S \times [0,T] \quad (19)$$

2. The potential of accelerations

$$\Theta = -(\dot{q} \mid p)_{R^n} - \Pi; \Pi = -\text{div}\dot{q}; \{q,p\} \in D_s \quad (20)$$

3. An impulse

$$p = \text{grad}\Psi; \quad q \in D_s; \quad \Psi = \Psi(q,t) \qquad (21)$$

4. The equation of the potential of accelerations

$$\frac{\partial \Psi}{\partial t} = \Theta; \quad \{q,p\} \in D_s \qquad (22)$$

5. The equation of an impulse

$$\frac{\partial p}{\partial t} = \text{grad}\Theta; \quad q \in D_s \qquad (23)$$

6. The equation of Entropy Time

$$\frac{dS}{dt} = \sigma_1; \quad t \in [0,T] \qquad (24)$$

7. The coupling equations

$$\Psi = H_q; \quad H_q = -\ln \eta.$$

It is useful to remind also, that except for the above mentioned symmetry, the symmetry of a duality of representation of the generalized coordinate exists here:

$$q = \begin{cases} q \in \Omega_q \\ q(t); t \in [0,T] \end{cases}.$$

The first component of this duality corresponds to the Euler's description, and the second – to Lagrange's one.

## 14. The matrix of density of an impulse

Development of the analytical theory of monitoring and algorithmization of the problem of the financial-economic monitoring is based on the representation of an impulse, given in the book [16]:

$$p = \Lambda q; q \in D_S; t \in [0, T]. \tag{25}$$

In this representation $\Lambda$ - the matrix of density of an impulse. In some cases the equation (25) underlies the transformation of the equation of an impulse (23) into matrix Riccati's equation. The matrix of density of an impulse opens ample opportunities for development of the mathematical theory of the financial-economic monitoring and effective algorithmization of the problem.

For example, in the well-known special case, having important practical sense for economy, the divergent invariant is defined by the following formula

$$\sigma = Sp\Lambda$$

For this case equation of Entropy Time becomes:

$$\frac{dS}{dt} = Sp\Lambda; t \in [0, T].$$

The important detail exists here; the matrix of density of an impulse has dual representation:

$$\Lambda = \begin{cases} \Lambda(q, t); & q \in D_S \\ \Lambda(t); & t \in [0, T]. \end{cases}$$

Let's address now to a problem of dimension of Euclidean spaces $R^n$. Let's remind that the most perfect model of econophysics - Heston's model - consists of two stochastic differential equations; numerous publications are devoted to its research, but the theory of this model is far from finalization. On the other hand, for economy the two-dimensional model has mostly the cognitive sense. Real tasks demand the greater dimensionality; and the next step hare is creation of three-dimensional model of the financial-economic monitoring.

A flow on Diffusion field allows the following representation for three-dimensional Euclidean spaces $R^3$:

$$\dot{q} = u; q \in D_S; u = \tilde{p}; t \in [0, T]. \tag{26}$$

In this representation $\tilde{p}$ is the auxiliary impulse.

In turn, it is possible to postulate the structure of the auxiliary impulse

$$\widetilde{p} = \widetilde{\Lambda}q; q \in D_S; t \in [0, T]$$ (27)

Here:

$\widetilde{\Lambda}$ - the matrix of density of the auxiliary impulse.

Now comes the key link: introduction of Helmholtz's matrix [17].

$$\widetilde{\Lambda} = \chi + \mu + \widetilde{\Omega}$$

$$\chi = \begin{Vmatrix} \chi_1 & 0 & 0 \\ 0 & \chi_2 & 0 \\ 0 & 0 & \chi_3 \end{Vmatrix}; \quad \mu = \begin{Vmatrix} 0 & \mu_1 & \mu_2 \\ \mu_1 & 0 & \mu_3 \\ \mu_2 & \mu_3 & 0 \end{Vmatrix}; \quad \widetilde{\Omega} = \begin{Vmatrix} 0 & -\omega_3 & \omega_2 \\ \omega_3 & 0 & -\omega_1 \\ -\omega_2 & \omega_1 & 0 \end{Vmatrix}$$ (28)

Here:

$\widetilde{\Lambda}$ - matrix of density of the auxiliary impulse;

$\chi$ - expansion matrix;

$\mu$ - shifting matrix;

$\widetilde{\Omega}$ - matrix of a rotor.

In terms of Helmholtz's matrix the equation of flow (26) will become:

$$\dot{q} = \widetilde{\Lambda}q; q \in D_S; t \in [0, T]$$

This flow has precise interpretation:
1. The matrix $\chi$ defines a spherical flow of expansion
2. The matrix $\mu$ defines a flow of shift
3. The matrix $\widetilde{\Omega}$ defines a flow of a rotor (whirl flow).

Here we come to the major conclusion: in three-dimensional Euclidean space the condition of the economic continuous medium is determined by two essences - dissipation and rotor.

It is clear, that in this task the divergent invariant will be

$$\sigma_1 = \mathrm{Sp}\chi.$$

The task of establishing parameters determining a condition of an object of the monitoring is important as well. The effective solution of this task is achieved on the basis of matrix Riccati's equation.

In Cauchy's task for matrix Riccati's equation the flow on Diffusion field is determined by four matrixes set above the field of real numbers:
1. *A* - the simple matrix of structure of object of monitoring
2. *B* - the symmetrical matrix of dissipation
3. *K* - the alternate matrix of a rotor
4. $\Lambda$ - the matrix of initial values of the matrix of density of an impulse.

## 15. Destruction and restoration of financial-economic structures

The important independent section of the theory of monitoring is formed with a problem of destruction and restoration of financial-economic structures.

The following problems of econophysics are situated at the basis of this unit:
1. Prediction of crisis situations in economy
2. Prevention of crisis situations in economy
3. Destruction of financial-economic structures
4. Post-crisis restoration of financial-economical structures.

In the entropy theory of monitoring these problems may be effectively researched. Classification of crisis conditions of econophysics as events lays in the basis of creation of the mathematical theory of destruction and restoration. The monograph "Entropy 2: Chaotic mechanics" is devoted to research of these events [17]. The characteristic feature of the book [17] is that events are realized on independent entropy manifold - extreme boundary layer (EBL). In econophysics the extreme boundary layer gives the basis for representation of entropy manifold as the universum

$$Э = Э_{-} \cup D_{s} \qquad (29)$$

In this universum the component $Э_{-}$ is the extreme boundary layer (EBL). EBL supports the principle of a limiting incorrectness of econophysics. The extreme boundary layer possesses symmetry - invariance of conditions. This symmetry implies autonomy of mathematical description of EBL. In this case, the components of universum $\{D_{s}, Э_{-}\}$ are studied independently at the first investigation phase; this independence reaches rather far, leading, finally, to independent branches of science or large theories. It happened with EBL: the manifold $Э_{-}$ has got independent object of research and has generated the independent section of the entropy conceptual model of Natural sciences - chaotic mechanics.

Symmetry - invariance of the description - does not imply absence of adjusting states of component of an universum in concrete problems. Here autonomy assumes independence of all set of potential states of EBL from the manifold $D_s$. But at the second investigation phase, at a choice of concrete realization there is concordance of conditions of the components of the universum Э.

The characteristic property of EBL is that it has small extension along the axis of astronomical time.

$$Э = Э_-, \forall t \in \sigma J, J = [0, T] \quad (30)$$

This property is effectively used in various problems and problems of the theory of monitoring.

In the present article we shall briefly discuss only two questions:
1. Simplification of the mathematical model
2. Classification of EBL.

The mathematical model of EBL is more simple comparing with the mathematical model of Diffusion field. The mathematical support of this model contains effective methods and ways of getting analytical solutions. The importance of classification of EBL is obliged to the fact of difference of its structure in different events. The existing classification of EBL is based on its measure [17].

**The statement**

*The measure of the extreme boundary layer is Jump* $[H_q]$ *of the structural entropy.*

The jump of the structural entropy at crossing EBL it is defined as follows

$$[H_q] = H_q^+ - H_q^-.$$

Here the index "-" is given for the structural entropy at the approach along the axis of time at the left (from the past), and the index "+" at the approach on the right (from the future).

Classification:
1. $[H_q] \neq 0$ - strong EBL

2. $[H_q] = 0$ - weak EBL.

Decomposition of a structure occurs in the strong EBL, and restoration of structure - in the weak EBL.

This property has underlain the formulation of a problem of restoration of structure:

$$[H_q] \neq 0 \quad \rightarrow \quad [H_q] = 0.$$

Let us note, that researches of processes of destruction and restoration have the long history. In this section of chaos the Moscow school [51-53] has nowadays the greatest popularity. S. V. Kurdjumov has introduced the special term «modes with aggravation» for this kind of events. The area of applicability of «modes with aggravation» is huge; more than two thousand works has been published on this theory. Intensive researches of processes of destruction and restoration are conducted also in the West. The review of the state of these researches is given in work [54].

Let us note in the conclusion that in the theory of monitoring the key fact has been established: *the rotor is of fundamental importance in a positive solution of the problem of post-crisis restoration of structure of an object of monitoring.*

Without the rotor the problem of restoration of structure has no solution.

The another important detail is that for smooth real functions the problem of restoration of structure has no solution also. The transition to distributions is necessary for its solution above a field of real numbers [55].